\documentclass[superscriptaddress,showpacs,aps,twocolumn,prb,floatfix]{revtex4}
\usepackage{amssymb}
\usepackage{amsmath}
\usepackage[dvips]{graphicx}
\usepackage{bm}

\begin{document}

\title{Calculation of the four-spin cyclic exchange in cuprates}
\author{A.~A.~Aligia}
\affiliation{Centro At\'{o}mico Bariloche and Instituto Balseiro, Comisi\'{o}n Nacional
de Energ\'{\i}a At\'{o}mica, CONICET, 8400 Bariloche, Argentina}
\email{aligia@cab.cnea.gov.ar}

\begin{abstract}
Starting from the three-band Hubbard model for the cuprates, we calculate
analytically the four-spin cyclic exchange in the limit of infinite on-site
Coulomb repulsion and zero O-O hopping $t_{pp}$ using two methods: i)
perturbation theory in $t_{pd}/\Delta$, where $t_{pd}$ is the Cu-O hopping
and $\Delta$ the Cu-O charge transfer energy and ii) exact solution of a Cu$_4$O$_4$ plaquette. 
The latter method coincides with the first to order
eight in $t_{pd}$ and permits to extend the results to $t_{pd}/\Delta$ of order
one. The results are relevant to recent experimental and theoretical
research that relate the splitting of certain spin excitations with $\Delta$
and the superconducting critical temperature.
\end{abstract}

\pacs{75.30.Et, 74.72.-h, 75.50.Ee, 71.70.Gm}
\maketitle



\section{Introduction}

\label{intro}

Several works have studied the influence of the energy and existence of apical oxygen (O) atoms
on the superconducting critical temperature $T_c$ of the cuprates.\cite{ohta,fei,saka} 
The absence of apical O atoms and larger separation between their on-site energy and that of the O atoms of
the superconducting CuO$_2$ planes increases $T_c$. The absence of negatively charged
apical O ions also renders less favorable to add holes to the neighboring copper (Cu)
ions reducing $\Delta$, the energy necessary to transfer a hole from the Cu atom to a nearest-neighbor 
O atom in the CuO$_2$ planes. In turn, decreasing $\Delta$ is expected to increase 
considerably the four-spin cyclic exchange $J_{4c}$ around the Cu$_4$O$_4$ 
square plaquettes, which has an important effect in the magnon dispersion \cite{coldea,peng}
and magnetic Raman  \cite{roger,honda} and infrared \cite{lore} spectrum for insulating cuprates.
In particular, the magnon
splitting at the Brillouin zone boundary $\Delta E_{\text{MBZB}}$ in simple spin models 
is proportional to $J_{4c}$.\cite{coldea}
Therefore, it is natural to expect that the magnitude of $J_{4c}$, measurable through
$\Delta E_{\text{MBZB}}$ gives information on the charge-transfer energy $\Delta$
and the expected $T_c$ in the cuprates, as shown by the recent work of Peng \textit{et al.}.\cite{peng}
The four-spin cyclic exchange also plays an important role in spin ladder cuprates.\cite{bre,matsu,calza}
It is also interesting to note that multiple spin exchange plays an essential role in the
thermodynamic properties of solid $^3$He in bulk \cite{godfrin} and in films.\cite{roger2}

Experimental evidence of the symmetry of holes in high-$T_c$ superconductors,\cite{nu,taki} 
and first-principles constrained-density-functional calculations,\cite{hyb2,hyb}
indicate that the appropriate model to describe the electronic structure of superconducting
CuO$_2$ planes is the three-band Hubbard model,\cite{eme,varma} which contains the 3d orbitals of the 
Cu atoms with $x^2-y^2$ symmetry and the 2p orbitals of the O atoms which point towards the Cu atoms. 
In terms of perturbation theory in the Cu-O hopping $t_{pd}$, the four-spin cyclic exchange $J_{4c}$
(although of order eight) is the non-trivial physical term of lowest order 
which does not involve double occupancy of holes at Cu or O sites. Therefore one expects that 
it pays an important role for small  Cu-O charge-transfer energy $\Delta$. 
However, to our knowledge, there is no calculation of $J_{4c}$ in the 
three-band Hubbard model.\cite{eme,varma}
Instead, a calculation is available in the one-band Hubbard model, where 
$J_{4c}$ is of fourth order in the hopping integral $t$.\cite{mac}
However, this result cannot be extended to the cuprates.  
While efficient low-energy reductions of the three-band to the one-band Hubbard model
exist which provide the values of $t \sim t_{pd}^2$ and the one-band on-site repulsion 
$U$,\cite{fei,hyb,schu,brin} they include terms which are at most of order $t_{pd}^4$, and therefore
some higher-order processes are lost if the one-band result for $J_{4c}$ is used.  

Recently, a numerical calculation of the magnon splitting $\Delta E_{\text{MBZB}}$
in a cluster of eight unit cells described by the three-band Hubbard model
has been reported.\cite{wang} It shows that $\Delta E_{\text{MBZB}}$ increases
as $\Delta$ decreases as expected, and the order of magnitude of the splitting agrees
with that measured in several compounds.\cite{peng} In any case, this is an expensive 
calculation and if a simpler calculation of $J_{4c}$ were available, this term 
could be introduced in spin models or successful generalized $t-J$ models,\cite{beli,sys,fei2,tcuo} 
which have a much smaller Hilbert space for the same cluster size and 
can be attacked with other techniques.\cite{tcuo,scba,lema,zeyher}

In this work we report on two analytical results for the four-spin cyclic exchange and the magnon splitting
starting from the three-band Hubbard model for infinite Cu ($U_d$) and O ($U_p$) on-site Coulomb repulsions
and zero O-O hopping $t_{pp}$: perturbation theory in the Cu-O hopping $t_{pd}$ up to order eight and exact
solution of a Cu$_4$O$_4$ plaquette. The latter is equivalent to include all higher order perturbation terms 
that are contained in this plaquette and leads to a considerable improvement of the results for small
$\Delta$.  For realistic and small values of $\Delta$ (of the order of 3.6 eV or smaller \cite{ohta,hyb2,hyb}), 
the assumption of infinite 
on-site repulsions is not essential. In fact it has been shown that our results are insensitive to the Coulomb
repulsion at the Cu sites $U_d$,\cite{wang} while perturbative processes involving O on-site repulsion $U_p$ 
do not contribute at order eight and involve large denominators at higher order.
The exact solution of the cluster permits to extend the validity of the results to smaller values of 
$\Delta > 2 t_{pd}$. For $\Delta < 2 t_{pd}$ other terms like the six-spin cyclic exchange affect the magnon splitting.

In Section \ref{model} we describe the three-band Hubbard model. 
In Section \ref{hspin} we explain the origin of the four-spin cyclic exchange. 
Section \ref{pertu} contains the result of perturbation theory in $t_{pd}/\Delta$ in lowest non-trivial order. 
In Section \ref{exact} we obtain the exact spectrum of a Cu$_4$O$_4$ cluster from which
a calculation of the four-spin cyclic exchange beyond perturbation theory is obtained, and discuss 
the range of validity and limitations of this calculation.
Section \ref{sum} contains a summary.

\section{Model} 
\label{model}

Our starting Hamiltonian corresponds to the three-band Hubbard model for cuprate superconductors \cite{eme,varma}

\begin{eqnarray}
H &=&\Delta \sum_{j\sigma }n_{j\sigma }-\sum_{\langle ij\rangle
}t_{pd}(p_{j\sigma }^{\dagger }d_{i\sigma }+\mathrm{H.c.})  \nonumber \\
&&+U_{d}\sum_{i}n_{i\uparrow }n_{i\downarrow }+U_{p}\sum_{j}n_{j\uparrow
}n_{j\downarrow }.  \label{ham}
\end{eqnarray}
Here $d_{i\sigma }^{\dagger }$ creates a hole with spin sigma in the 3d orbital of Cu at site $i$ with symmetry
$x^2-y^2$. Similarly $p_{j\sigma }^{\dagger }$ creates a hole in the O orbital at site $j$ which is directed to 
the nearest-neighbor Cu atoms (they are usually called $p_\sigma$ orbitals). The relevance of these orbitals over
the rest is justified by experimental evidence \cite{nu,taki} and first-principles calculations.\cite{hyb2,hyb} 
The hole number operators are  $n_{i\sigma }=d_{i\sigma }^{\dagger }d_{i\sigma }$ and 
$n_{j\sigma }=p_{j\sigma }^{\dagger }p_{j\sigma }$.

In this work we neglect the O-O hopping $t_{pp}$ and take $U_d, U_p \rightarrow \infty$. The consequences of these 
approximations are discussed in Section \ref{exact}. 

\section{Form of the four-spin cyclic exchange}
\label{hspin}

\begin{figure}[h]
\begin{center}
\includegraphics[width=\columnwidth]{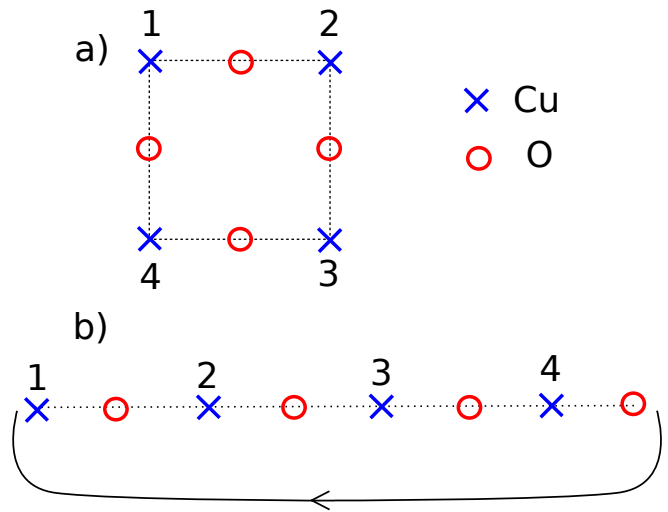}
\end{center}
\caption{(Color online) a) basic plaquette to calculate the four-spin cyclic exchange. 
b) equivalent linear chain with periodic boundary conditions.}
\label{plaq}
\end{figure}

The cyclic exchange acting on the four spins at the Cu sites in a square
plaquette as represented in Fig. \ref{plaq} is

\begin{equation}
H_{4c}=J_{4c}\left( C_{4}+C_{4}^{-1}\right) ,  \label{4c}
\end{equation}%
where $C_{4}$ is a cyclic permutation of the position of the four spins ($%
C_{4}|\sigma _{1}\sigma _{2}\sigma _{3}\sigma _{4}\rangle =|\sigma
_{4}\sigma _{1}\sigma _{2}\sigma _{3}\rangle $). The extension to the whole
lattice of Cu sites is straightforward \cite{mac}. Expressing  $C_{4}$ as a
product of transpositions one has

\begin{equation}
C_{4}+C_{4}^{-1}=P_{12}P_{23}P_{34}+P_{34}P_{23}P_{12}.  \label{suma}
\end{equation}%
Noting that $P_{il}=(1/2+2\mathbf{S}_{i}\cdot \mathbf{S}_{l})$ transposes
the spins $i$ and $l$ and performing the products of repeated Pauli matrices
in Eq. (\ref{suma}) one obtains after some algebra

\begin{eqnarray}
H_{4c} &=&J_{4c}[4\{(\mathbf{S}_{1}\cdot \mathbf{S}_{2})(\mathbf{S}_{3}\cdot 
\mathbf{S}_{4})+(\mathbf{S}_{1}\cdot \mathbf{S}_{4})(\mathbf{S}_{2}\cdot 
\mathbf{S}_{3})  \nonumber \\
&&-(\mathbf{S}_{1}\cdot \mathbf{S}_{4})(\mathbf{S}_{2}\cdot \mathbf{S}%
_{3})\}+\sum_{i<l}\mathbf{S}_{i}\cdot \mathbf{S}_{l}+1/4].  \label{4c2}
\end{eqnarray}%
In some papers only the four-spin term is included explicitly  (with the
prefactor $4J_{4c}$ denoted as $J_{ring}$ in Ref. \onlinecite{calza} or $J_{c}$ in
Refs. \onlinecite{coldea,peng}), leaving the other terms as corrections to
nearest-neighbor and next-nearest-neighbor exchange. In the one-band Hubbard model
fourth-order perturbation theory leads to $J_{4c}=20t^{4}/U$, $t$ being the
hopping integral and $U$ the on-site Coulomb repulsion.\cite{mac}

\section{Perturbation theory in the Cu-O hopping}

\label{pertu}

We split the Hamiltonian into the perturbation $H_{t}=-\sum_{\langle
ij\rangle }t_{pd}(p_{j\sigma }^{\dagger }d_{i\sigma }+\mathrm{H.c.})$ and $%
H_{0}=H-H_{t}$. The degenerate ground state of $H_{0}$ consists of a hole
occupying each Cu site and has energy $E_{f}^{0}=0$.  The processes of
lowest order in the Cu-O hopping $t_{pd}$ that contribute to $H_{4c}$ are of
order $t_{pd}^{8}$ and involve each of the plaquettes of the form represented
in Fig. \ref{plaq}. We can restrict the analysis to one of them and label
the sites as in Fig. \ref{plaq}. The perturbation processes that  mix a
state $|i\rangle =d_{1\sigma _{1}}^{\dagger }d_{2\sigma _{2}}^{\dagger
}d_{3\sigma _{3}}^{\dagger }d_{4\sigma _{4}}^{\dagger }|0\rangle $ (with
spin configuration $|\sigma _{1}\sigma _{2}\sigma _{3}\sigma _{4}\rangle $)
with the final state -$|f\rangle =-d_{2\sigma _{1}}^{\dagger }d_{3\sigma
_{2}}^{\dagger }d_{4\sigma _{3}}^{\dagger }d_{1\sigma _{4}}^{\dagger
}|0\rangle $ = $d_{1\sigma _{4}}^{\dagger }d_{2\sigma _{1}}^{\dagger
}d_{3\sigma _{2}}^{\dagger }d_{4\sigma _{3}}^{\dagger }|0\rangle $ (with
spin configuration $C_{4}|i\rangle =|\sigma _{4}\sigma _{1}\sigma _{2}\sigma
_{3}\rangle $) involves two hoppings of each electron in the clockwise
direction (Fig. \ref{plaq}), Adding all contributions (or the analogous ones
in the anticlockwise direction) one can obtain $J_{4c}$. From standard
degenerate perturbation theory one has

\begin{equation}
J_{4c}=\frac{\langle i|H_{t}|e_{1}\rangle \prod\limits_{j=1}^{6}\langle
e_{j}|H_{t}|e_{j+1}\rangle \langle e_{7}|H_{t}|f\rangle }
{\prod\limits_{j=1}^{7}(-E_{j}^{0})},  \label{jcper}
\end{equation}%
where $|e_{j}\rangle ,E_{j}^{0}$ denote the seven intermediate eigenstates
of $H_{0}$ and their energies. None of the intermediate states involve
double occupancy at an O site. Since we take $U_{d}\rightarrow +\infty $, we
neglect intermediate states with double occupancy at any Cu site. With the
help of a computer program we have obtained the remaining 1088 processes and
added the corresponding contributions to  $J_{c}$. The result is

\begin{equation}
J_{4c}=20\frac{t_{pd}^{8}}{\Delta^7}.  \label{jc}
\end{equation}

\section{Exact results for Cu$_{4}$O$_{4}$}

\label{exact}

The exact solution of the plaquette represented in Fig. \ref{plaq} permits
to extend the perturbation result to the covalent region in which $t_{pd}$
is not much smaller than $\Delta $. The procedure, similar to that followed
in other works \cite{sf,calza} is to fit the lowest energy levels of  $H$
with all those of $H_{4c}$ in the Cu$_{4}$O$_{4}$ cluster. It is equivalent to 
include all perturbation terms that are contained in the cluster.
Fortunately, as we shall show, the form Eq. (\ref{4c}) [or the equivalent one 
Eq. (\ref{4c2})] still describes the corresponding contribution to the effective 
Hamiltonian in a wide range of values of $\Delta$.

The spectrum of $H_{4c}$ is easy to obtain. Starting from any of the 16
states \  $|e\rangle =|\sigma _{1}\sigma _{2}\sigma _{3}\sigma _{4}\rangle $, 
one constructs eigenstates of $C_{4}$

\begin{equation}
|k,e\rangle =N\sum\limits_{j=0}^{3}(e^{-ik}C_{4})^{j}|e\rangle ,  \label{ke}
\end{equation}%
such that  $C_{4}|k,e\rangle =e^{ik}|k,e\rangle $. where $N$ is a
normalization factor and $k=m\pi /2$, where $m$ is an integer with
non-equivalent values 0, $\pm 1$, 2. Using Eq. (\ref{4c}) the energies become 

\begin{equation}
E_{k}=2J_{4c}\cos k.  \label{ek}
\end{equation}%
They only depend on $k$. An analysis of the other quantum numbers shows that
the ground state, which has wave vector $k=\pi $ and energy $E_{\pi }=-2J_{4c}$, 
contains a singlet and a triplet. The first excited states with $k=\pm \pi /2$ and $E_{k}=0$
are two triplets and the remaining six states with $k=0$ and $E_{0}=2J_{4c}$
are the quadruplet and the other singlet.

To solve the fermion  multiband model $H$ in the cluster we follow a simple
extension of the elegant procedure of Caspers and Ilske for the exact
solution of the Hubbard chain with infinite $U$.\cite{casp} Here we describe
the main idea. The details can be found in Ref. \onlinecite{casp}. 
The Cu$_{4}$O$_{4}$ cluster is equivalent to a linear chain with periodic boundary
conditions (see  Fig. \ref{plaq}). Imagine for the moment that the Cu-O hopping
between the last atom in the chain and the first one is set to zero, leaving
a chain with open boundary conditions. Then the holes hop between different
atoms in the chain but the order of the four spins ($\sigma _{1}\sigma
_{2}\sigma _{3}\sigma _{4}$)  is kept, since the infinite Coulomb repulsion
at each site does not allow to exchange spins. Furthermore, there is one to
one correspondence between any state of the system  and that obtained
replacing the spin configuration ($\sigma _{1}\sigma _{2}\sigma _{3}\sigma
_{4}$) by ($\uparrow \uparrow \uparrow \uparrow $).  The Hamiltonian matrix
in both spaces have the same form and since the problem in the latter
subspace is a spinless fermion problem, it can be solved trivially, and the
mapping provides a solution to the original problem. 

When the hopping
between the first and the last atom is restored, one is faced with the difficulty 
that when the
hole of the last atom hopes to the first one, the spin configuration 
($\sigma _{1}\sigma _{2}\sigma _{3}\sigma _{4}$) is changed to ($\sigma
_{4}\sigma _{1}\sigma _{2}\sigma _{3}$) = $C_{4}$($\sigma _{1}\sigma
_{2}\sigma _{3}\sigma _{4}$), where $C_{4}$ is a cyclic permutation of the
spin configuration\ (without affecting the charge distribution) and in
general, the above one to one correspondence is lost. However, for any
charge configuration, one can construct eigenstates of $C_{4}$ [similar to
Eq. (\ref{ke})]. It is easy to realize that for these eigenstates, the
mapping to the spinless Hamiltonian is still possible but the hopping from the
last atom to the first one becomes multiplied by $e^{ik}$, the eigenvalue of 
$C_{4}$. Clearly the reverse process has a factor $e^{-ik}.$ Then, the
problem becomes equivalent to spinless fermions under a magnetic flux.
While the original argument was developed for the Hubbard model, clearly it
is still valid if the on-site energies of the different sites differ.

For the solution of the equivalent spinless problem, it is convenient to
distribute the phase $e^{\pm ik}$ of the hopping term equally in all the eight Cu-O links by a gauge
transformation, so that translation symmetry with periodic boundary conditions is
restored in the equivalent spinless model and the hopping term takes the form
$H_{t}=-\sum_{i \delta }t_{pd}(e^{ik \delta /4}p_{i + \delta}^{\dagger }d_{i}+\mathrm{H.c.})$,
where $\delta = \pm 1/2$ and 
$p_{i+1/2}^{\dagger }$ ($p_{i-1/2}^{\dagger }$) creates a spinless O hole half a lattice
parameter at the right (left) of Cu site $i$. 

The spinless problem is solved as
usual in a basis of Cu and O one-particle states with charge wave vector 
$q=n\pi /2$ with $n=0$, $\pm 1$, 2. There are eight different one-particle eigenvalues, 
two for each $q$. Since
the system has four holes, one has to fill the four one-particle states of
lowest energy to obtain the low-energy spectrum that maps onto  $H_{4c}$. 
It turns out that for all $q$ only the lowest one-particle
state is occupied. The resulting many-body energies that map onto those of 
$H_{4c}$ for each spin wave vector $k$ become

\begin{equation}
E_{k}^{H}=2\Delta -\sum\limits_{n=-1}^{2}\sqrt{\left( \frac{\Delta }{2}\right) ^{2}
+4t_{pd}^{2}\cos ^{2}\left( \frac{n\pi }{4}+\frac{k}{8}\right) }.
\label{ekh}
\end{equation}

For small $t_{pd}$ this result can be expanded in powers of $t_{pd}/\Delta $. 
Only even powers of $t_{pd}$ enter. Although it is not apparent from Eq. (\ref{ekh}), the terms of second, 
fourth and sixth order give a result
independent of $k$, so that the first non trivial term is of order eight.
Except for an irrelevant constant, this expansion to order eight gives  $E_{k}^{H}=E_{k}$, 
where $E_{k}$ is the result obtained previously by perturbation theory [Eqs. (\ref{jc}) and (\ref{ek})]. 
Thus, as expected, for small  $t_{pd}/\Delta $, Eq. (\ref{ekh}) reproduces the perturbative 
result up to order eight in $t_{pd}$, but
with much less effort. Furthermore defining 

\begin{equation}
J_{4c}=\frac{E_{0}^{H}-E_{\pi }^{H}}{4},  \label{jcr}
\end{equation}%
permits to extend the validity of the low-energy Hamiltonian $H_{4c}$ to
larger values of $t_{pd}/\Delta $. 

\begin{figure}[h]
\begin{center}
\includegraphics[width=\columnwidth]{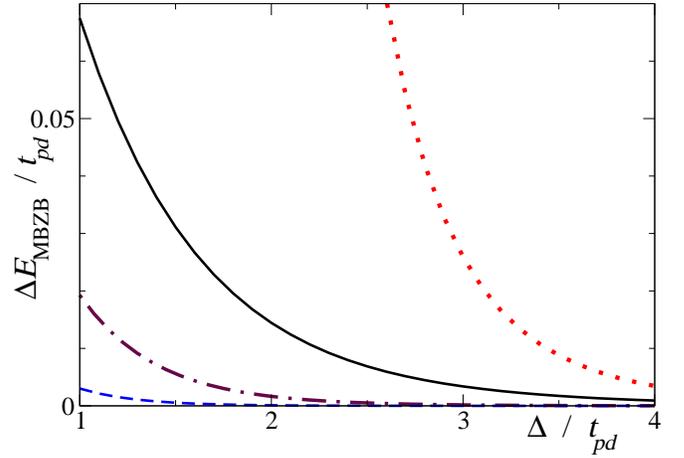}
\end{center}
\caption{(Color online) Magnon splitting (full black line) 
and perturbative result (dotted red line) as a function of $\Delta$.
Also shown for comparison (see text)  are $0.706D$ (dashed blue line) 
and $0.706S_6$ (dashed-dotted brown line).}
\label{cyc}
\end{figure}

Note that for a general Hamiltonian
with spin SU(2)  symmetry  the 16 low-energy eigenstates for 4 Cu spins in a
square lattice are expected to be split in five different eigenvalues 
(corresponding to either different wave vectors $k$ or different total spin), while
the eigenstates of $H_{4c}$ split only in three different
energies according to the value of $k$. This property is retained by the 
low-energy eigenstates of the full Hamiltonian $H$ in the Cu$_{4}$O$_{4}$ cluster, 
indicating that $H_{4c}$ continues to be a good representation of 
this low-energy subspace. However, a shortcoming of $H_{4c}$ 
in reproducing the eigenvalues of $H$ for large $t_{pd}/\Delta $ is that
in the latter the difference 

\begin{equation}
D=(E_{0}^{H}+E_{\pi }^{H})/2-E_{\pi /2}^{H},  \label{dif}
\end{equation}
is larger than zero
while the corresponding difference vanishes in $H_{4c}$. However, $D$ 
is very small for $\Delta > t_{pd}$ (see dashed line in Fig. \ref{cyc}).
The first correction to $D$ in powers of $t_{pd}$ [obtained evaluating numerically Eq. (\ref{ekh})]
is $6864 t_{pd}^{16}/ \Delta^{15}$.

A magnitude of interest in recent works \cite{peng,wang} is the magnon
splitting at the Brillouin zone boundary which for the spin model is given
by \cite{coldea,peng}

\begin{equation}
\Delta E_{\text{MBZB}}=\frac{12}{5}Z_{c}J_{4c},  \label{spli}
\end{equation}
where $Z_c=1.18$ is a renormalization factor accounting for quantum
fluctuations.\cite{singh}
This quantity using $J_{4c}$ given by Eq. (\ref{jcr}) is represented 
by the full line in Fig. \ref{cyc} as a function of $\Delta$.
The corresponding perturbative result using $J_{4c}$ 
given by Eq. (\ref{jc})  
(dotted line in in Fig. \ref{cyc}) largely
overestimates the splitting for realistic values of $\Delta$ (smaller than $4t_{pd}$).
Taking into account that estimated values of $\Delta$ ($t_{pd}$) are in the range 1.3-3.6 (1.1-1.6) eV,\cite{hyb2}
our results for $\Delta E_{\text{MBZB}}$ due to $J_{4c}$ are approximately in the range 2-110 meV, 
which can be compared with the values between 30 and 150 meV measured in four parent compounds
of high-$T_c$ superconductors.\cite{peng}
Our result is qualitatively similar to the splitting calculated in small clusters 
described by the three-band Hubbard model [Fig. 2(b) in Ref. \onlinecite{wang}].
The result in that work is larger due to the inclusion of the O-O hopping
$t_{pp}$ which we have neglected here. We believe that to a first approximation,
the main effect of $t_{pp}$ is similar to decrease $\Delta$ by $1.46 |t_{pp}|$,
as suggested by a change of basis to O orbitals centered on Cu sites.\cite{beli}

For small $\Delta$, other perturbative processes of high order in $t_{pd}/\Delta$,
which are not contained in the Cu$_4$O$_4$ plaquette, might become important.
The dominant one is probably the six-spin cyclic exchange, which can be calculated as above
in a Cu$_6$O$_6$ cluster. 
In Fig. \ref{cyc} we also show the energy difference given by Eq. (\ref{dif}) and the magnitude of 
the six-spin cyclic exchange $S_6=E_0-E_\pi$ calculated as above in a Cu$_6$O$_6$
ring rescaled by the factor $f=Z_c \times 3/5$ to evaluate their relative magnitude 
compared to $\Delta E_{\text{MBZB}}$ (neglecting $S_6$, $\Delta E_{\text{MBZB}}=f (E_0^H-E_\pi^H)$). 
The conclusion of this comparison is that while 
the effective Hamiltonian $H_{4c}$ represents accurately the effects of the four-spin cyclic 
exchange for $\Delta > t_{pd}$, other terms, like the six-spin cyclic exchange become 
important for $\Delta < 2 t_{tpd}$ and affect $\Delta E_{\text{MBZB}}$.

\section{Summary}
In summary, for the three-band Hubbard model with infinite Coulomb repulsions
$U_d$ and $U_p$, O-O hopping $t_{pp}=0$, and $\Delta \geq 2 t_{pd}$, 
the magnon splitting at the Brillouin zone boundary is accurately described by the analytical 
expressions Eqs. (\ref{spli}), (\ref{jcr}) and (\ref{ekh}). Finite Coulomb repulsions are expected to have a
very minor effect. Inclusion of $t_{pp}$ increases the splitting. We expect that
the main effect of increasing $t_{pp}$ is equivalent to a decrease in $\Delta$.
For $\Delta < 2 t_{pd}$, the six-spin cyclic exchange (which can be calculated following
the lines of this work) also becomes important and affects the magnon splitting,
The analytical expressions permit a rapid estimation of the magnon splitting (a lower bound 
for sizable $t_{pp}$). Conversely, given a magnon splitting measured experimentally
one can infer the magnitude of $\Delta$, and from it, one might have a qualitative idea of 
the expected superconducting critical temperature.

\label{sum}

\section*{Acknowledgments}

This work was sponsored by PIP 112-201501-00506 of CONICET and PICT
2013-1045 of the ANPCyT.

\end{document}